# The quantum phase slip phenomenon in superconducting nanowires with high-impedance environment

K. Yu. Arutyunov • J. S. Lehtinen • T. Rantala

**Abstract**. Quantum phase slip (QPS) is the particular manifestation of quantum fluctuations of the order parameter of a current-biased quasi-1D superconductor. The QPS event(s) can be considered a dynamic equivalent of tunneling through conventional Josephson junction containing static in space and time weak link(s). At low temperatures $T \ll T_c$ the QPS effect leads to finite resistivity of narrow superconducting channels and suppresses persistent currents in tiny nanorings. Here we demonstrate that the quantum tunneling of phase may result in Coulomb blockade: superconducting nanowire, imbedded in high-Ohmic environment, below a certain bias voltage behaves as an insulator.

**Keywords** Nanoscale superconductivity • Quantum fluctuations

K. Yu. Arutyunov ()
National Research University Higher School of Economics, Moscow Institute of Electronics and Mathematics 101000, Moscow, Russia
P.L. Kapitza Institute for Physical Problems RAS, Moscow, 119334, Russia
E-mail: karutyunov@hse.ru

J. S. Lehtinen
VTT Technical Research Centre of Finland Ltd., Centre for Metrology MIKES, P.O. Box 1000, FI-02044 VTT, Finland

T. Rantala
Department of Physics, University of Jyvaskyla, PB 35, FI-40014 Jyvaskyla, Finland

## 1 Introduction

Modern nanotechnology enables reproducible fabrication of artificial systems with dimensions approaching ~10 nm scales. In superconductors the characteristic dimensionality is set by the coherence length $\xi$. For conventional lift-off fabricated polycrystalline aluminium or titanium nanostructures the coherence length can be made as large as $\xi \sim 100$ nm. Hence, it is quite realistic to fabricate and study quasi-one dimensional (1D) superconducting systems [1]. It has been demonstrated that in such objects quantum fluctuations of the phase $\varphi$ of the complex order parameter $\Delta = |\Delta|e^{i\varphi}$ can suppress zero resistivity at temperatures well below the critical point $T \ll T_c$ [2-7] and dramatically modify the current-phase relation and the magnitude of persistent currents in narrow superconducting nanorings [8,9].

It has been pointed out [10] that the Hamiltonians describing a Josephson junction (JJ)

$$\widehat{H}_{JJ} = E_C \hat{q}^2 - E_J \cos(\hat{\varphi}) + \widehat{H}_{COUP} + \widehat{H}_{ENV} \quad (1)$$

and a short superconducting nanowire in the regime of QPS, which correspondingly can be called the *quantum phase slip junction* (QPSJ)

$$\widehat{H}_{QPSJ} = \frac{E_L}{(2\pi)^2}\hat{\varphi}^2 - E_{QPS}\cos(2\pi\hat{q}) + \widehat{H}_{COUP} + \widehat{H}_{ENV} \quad (2)$$

are identical with accuracy of substitution $E_C \leftrightarrow E_L$, $E_J \leftrightarrow E_{QPS}$ and $\varphi \leftrightarrow \pi q/2e$, where $E_C$, $E_L$, $E_J$, $E_{QPS}$ are the energies associated with charge, inductance, Josephson and QPS couplings, $q$ is the quasicharge and $2e$ is the charge of a Cooper pair. $\widehat{H}_{COUP}$ and $\widehat{H}_{ENV}$ are the coupling and environmental Hamiltonians, which can be the same for a JJ and a QPSJ. The identity of the Hamiltonians (1) and (2) reflects the fundamental quantum duality of these two systems.

The energy $E_{QPS}$ is related to the rate of QPS events and it exponentially depends on the cross section $\sigma$ of the superconducting channel and material parameters [1,3]

$$E_{QPS} \sim \exp\left(-a\frac{\sigma T_c^{1/2}}{\rho_N}\right) \quad (3)$$

, where $\rho_N$ is the normal state resistivity, $T_c$ is the critical temperature and $a$ is the constant of proper dimensionality. Hence, narrow nanowires of low-$T_c$ superconductors with high resistivity in normal state should be the first choice for the QPS effect study. Note that exponential dependence (3) enables easier, compared to conventional JJs, fabrication of QPSJs with almost arbitrary relation between the characteristic energies $E_C$, $E_L$ and $E_{QPS}$.

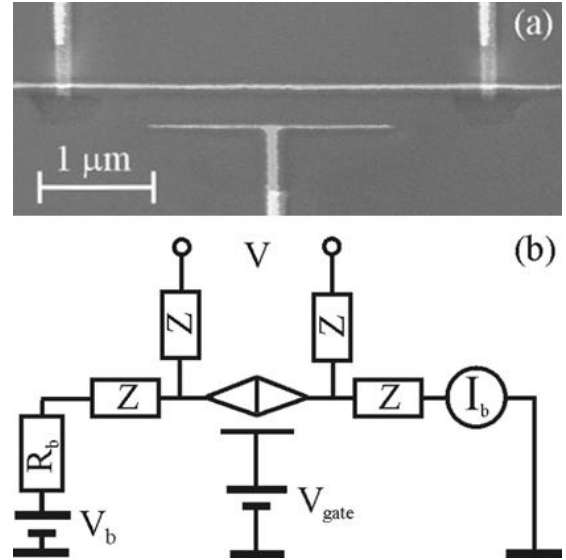

**Fig. 1** (a) SEM image of the central part of a typical nanostructure. (b) Schematics of the equivalent circuit. Central rhomb stands for the QPS nanowire. Rectangles marked 'Z' correspond to high-impedance on-chip elements.

Physics of a conventional JJ differs depending on how the system is biased – voltage or current – corresponding to phase $\varphi$ or charge $q$ acting as a quasiclassical variable. Quantum dynamics of a current-biased JJ with large charging energy $E_C > E_J$ can be described by equations similar to motion of a particle in a periodic potential [11,12]. The corresponding Bloch oscillations have been observed in ultra-small JJs [13] and QPSJs [14-16]. The objective of this paper is to study the current-voltage characteristics of current-biased thin superconducting channels in the regime of QPS.

## 2 Experiment

Earlier experiments on superconducting titanium nanowires with effective diameter $\sigma^{1/2} \lesssim 45$ nm probed by low-Ohmic (superconducting) electrodes have demonstrated broad R(T) dependencies, associated with manifestation of the QPS effect [7]. In this work we study titanium nanostructures of similar dimensions, but, contrary to those studies, the new structures are imbedded in high impedance environment (Fig. 1). High-Ohmic resistors (up to 10 M$\Omega$) or/and chains of Josephson junctions were used as on-chip high impedance elements (marked 'Z' in Fig.1) enabling current bias mode. Low energy directional ion milling was used to reduce the critical dimensions of the lift-off fabricated nanostructures down to sub-20 nm scales [17]. The capacitively coupled central T-shaped electrode was used as an electrostatic gate. The multi stage RLC filters (not shown in Fig.1) enabled reduction of the electron temperature down to ~ 30 mK at the base temperature of our $^3$He$^4$He dilution refrigerator ~ 17 mK [18,19]. All input/output lines between the battery powered analogue front-end electronics and the rest of the measuring set-up (PC, DMMs, lock-in) were carefully filtered passing through the walls of the EM shielded room.

## 3 Results and discussion

The I-V characteristics of *Ti* QPS nanowires with low-impedance probes, have been studied earlier [7,20]. In those experiments sufficiently thick samples with $\sigma^{1/2} > 45$ nm demonstrated conventional I-V dependencies with well-defined critical current. The I-Vs of narrower structures revealed some nonlinearities, which could be associated with 'residual' critical current, while no true zero voltage stage has been detected in the thinnest nanowires even at temperatures $T \ll T_c$.

The I-V dependencies of the structures with high-impedance probes, studied in present work, are qualitatively different. All structures with sufficiently narrow nanowire and the impedance of the probes above several hundred k$\Omega$, revealed counterintuitive behavior for a superconductor: insulating state below a certain critical voltage (Fig. 2). Following formalism, developed for the dual quantum system - Josephson junction - one can conclude that such a behavior is quite expected [11-13]. In the current biased mode the quasicharge $q$ behaves as a quasiclassical variable, making the conjugated variable - the phase $\varphi$ - undefined, thus enabling strong phase fluctuations. The insulating state - the Coulomb blockade - is the result of the coherent superposition of the QPSs in the nanowire [10]. Note that the current biased mode, provided by the high impedance on-chip probes and the ballast resistor (marked as Z and $R_b$ in Fig. 1) is possible only at a finite current. In the insulating state the impedance of the nanowire tends to infinity. Thus entering the Coulomb blockade the measuring circuit switches from current to voltage biased modes. This peculiarity leads to technical problem in recording such I-Vs. Given the very high impedance of the circuit approaching the insulating state, the corresponding time constants (taking into consideration the mandatory capacitance of the RLC filters) reach tens of seconds. Thus the rate of the sweep dramatically affects the shape of the I-V dependence. A typical example of relatively quickly recorded I-V is shown in Fig. 2. Much more slow sweep reveals back-bending regions ('Bloch noses') studied in Ref. 14. In this work, in addition to DC recorded I-Vs, we also used modulation technique: the slowly ramped DC bias current $I_b$ has been modulated by small AC current. The signal, measured by lock-in amplifier, is proportional to the differential resistance dV/dI (Fig. 2, inset). One can clearly see that within the Coulomb blockade the differential resistance dramatically increases.

The differential resistance at zero bias depends on potential of the gate electrode $V_{gate}$ (Fig.3). The corresponding periodic modulation is the well-known effect and is used for operation of a single electron transistor or a Copper pair box [21].

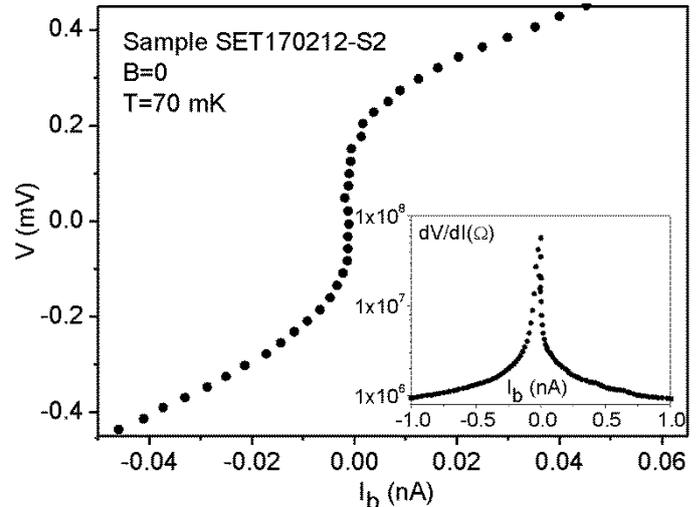

**Fig. 2** Zoom of the central part of relatively quickly recorded two-probe measured I-V characteristic of titanium nanowire with effective diameter $\sigma^{1/2}$=35±3 nm. Inset shows larger scale dependence of the four-probe measured differential resistance dV/dI vs. bias current recorded using modulation technique.

The width of the Coulomb gap should correlate with the rate of QPS events $\delta V_C \sim E_{QPS}/2e$ and thus, following Eq. 3, should exponentially depend on cross section of the nanowire $\sigma$. The observation correlates well with our findings: no traces of Coulomb effects were observed in samples with effective diameter $\sigma^{1/2} \gtrsim 50$ nm. Note that nanowires with similar 'large' dimensions contacted with low impedance probes have revealed no traces of QPS effects: rather sharp R(T) dependencies, well defined critical current and zero resistive state [7,20].

In the low temperature limit $T \ll T_c$ the characteristic energy $E_{QPS}$ weakly depends on temperature [1,3]. This feature is supported by our data: the magnitude of the differential resistance modulation by the gate potential dV/dI($V_{gate}$) is almost insensitive to temperature in the low temperature limit $T \ll T_c$, decreases at high temperatures and completely disappears above ~ 500 mK, which roughly correlates with the critical temperature of our 'dirty' titanium (Fig. 3). The observation is very important for association of the experimentally observed Coulomb blockade with the QPS effect, and not with unintentionally formed weak links. If the latter case, one would observe dV/dI($V_{gate}$) modulation at $T>T_c$ attributed to single electron phenomena. Elementary estimation of the minimal charging energy $E_C$, corresponding to maximal capacitance of a weak link (e.g. insulating barrier) of a $\sigma^{1/2}$=35 nm nanowire, gives $E_C/k_B \approx 7$ K. Hence, single electron effects should manifest themselves in much larger values of the Coulomb gap $\delta V_C$ and 'survive' up to comparable temperatures, which is clearly not the case (Fig. 3).

At low temperatures $T \ll T_c$ gate modulation of the differential resistance within the Coulomb blockade is suppressed by application of sufficiently strong magnetic field ~ 3 T. The observation also supports the conclusion that the effect is somehow related to dynamic superconducting phenomena and not to unintentionally formed static tunnel barriers. Otherwise in normal state the system should exhibit single electron gate modulation up to much higher magnetic fields [22].

It should be noted that dependence dV/dI($V_{gate}$) has larger magnitude for 'even' maxima corresponding to overlapping of 1$e$ and 2$e$ (Cooper pair) singularities (Fig. 3). One may conclude that though at T<<$T_c$ the system is in superconducting state, the single charge effects do present. The origin of this anomaly is not clear, and presumably (?) is related to contribution of nonequilibrium quasiparticles, created at each phase slip event [23].

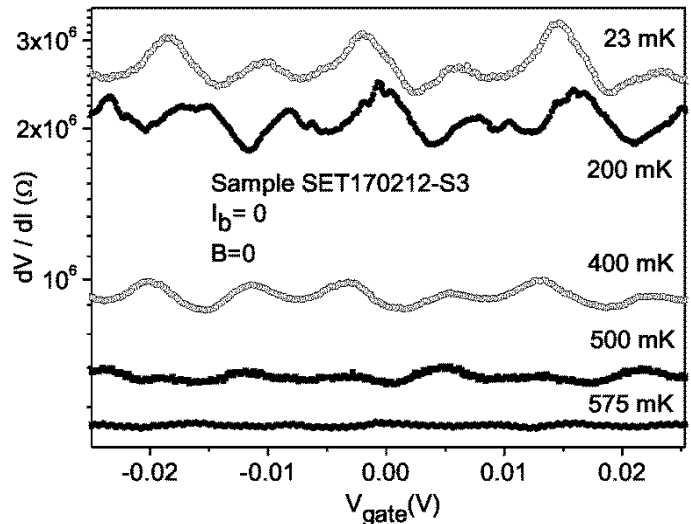

**Fig. 3** Gate modulation of the differential resistance within the Coulomb gap at various temperatures for titanium nanowire with effective diameter $\sigma^{1/2}$=43±3 nm. Differential resistance dV/dI is measured at zero bias current $I_b$=0 using modulation technique.

**Conclusions**

We have studied the I-V characteristics of thin current-biased titanium nanowires probed by high impedance contacts. Below a certain critical voltage the system demonstrates Coulomb blockade associated with coherent superposition of quantum phase slips. The effect disappears above critical temperature and/or magnetic field supporting the relation of the phenomena to superconductivity and not to unintentionally formed tunnel junctions. Analogous behavior has been observed in ultra-small current-biased Josephson junctions [13]. The similarity originates from the fundamental quantum duality between these two systems [10].

**Acknowledgments**

The authors would like to acknowledge D. Averin, P. Hakkonen, T. Heikkilä, L. Kuzmin, Yu. Nazarov, Yu. Pashkin, J. Pekola, A. Zaikin, and A. Zorin for valuable discussions.